\documentclass{aa}

\usepackage{graphicx} % Required for inserting images
\usepackage{subfigure}
\usepackage{caption}
\usepackage{float}
\usepackage{txfonts}
\usepackage{siunitx}
\usepackage{xspace}
\usepackage{float}
\usepackage[switch]{lineno}
\usepackage[colorlinks,allcolors=blue]{hyperref}
\usepackage{multicol}
\usepackage{multirow}
\usepackage{makecell}

\makeatletter
\renewcommand*\aa@pageof{, page \thepage{} of \pageref*{LastPage}}
\makeatother

\begin{document}

\title{Event types in H.E.S.S.: A combined analysis for different telescope types and energy ranges}

\titlerunning{Event types in H.E.S.S.}

\author{
  Rodrigo Guedes Lang\inst{\ref{ECAP}} \and
  Tim Unbehaun\inst{\ref{ECAP}} \and
  Lars Mohrmann\inst{\ref{MPIK}} \and
  Simon Steinmassl\inst{\ref{MPIK}} \and
  Jim Hinton\inst{\ref{MPIK}} \and 
  Stefan Funk\inst{\ref{ECAP}}
}

\institute{
Friedrich-Alexander-Universit\"at Erlangen-N\"urnberg, Erlangen Centre for Astroparticle Physics, Nikolaus-Fiebiger-Str. 2, 91058 Erlangen, Germany\\ \email{rodrigo.lang@fau.de} \label{ECAP} \and
Max-Planck-Institut f\"ur Kernphysik, Saupfercheckweg 1, 69117 Heidelberg, Germany\\ \label{MPIK} 
}

\abstract{Imaging atmospheric Cherenkov telescopes (IACTs) are the main technique for detecting gamma rays with energies between tens of GeV and hundreds of TeV. One of them, the High Energy Stereoscopic System (H.E.S.S.), has pioneered the use of different telescope types to achieve the broadest possible energy range. A large telesecope with a diameter of $28 \, \rm{m}$ (CT5) is used in monoscopic mode to access the lowest energies ($E \gtrsim 30 \, \rm{GeV}$), while the four smaller telesceopes with diameters of $12 \, \rm{m}$ (CT1 to CT4) are used in stereoscopic mode to study energies between 150 GeV and 100 TeV. Nevertheless, a combination of the two telescope types and trigger strategies has proven to be challenging. We propose an analysis based on event types capable of exploiting both telescope types, trigger strategies, and the whole energy range of the experiment for the first time. Because monoscopic and stereoscopic reconstructions are very different, the types are defined based on the Hillas parameters of individual events, resulting in three types (\texttt{Type M}, \texttt{Type B}, and \texttt{Type A}), each of which dominates a different energy range. Further improvements to the gamma-hadron separation and to the energy and angular reconstruction are introduced as well. The performances of the new analysis configurations are compared to the standard configurations in the H.E.S.S. Analysis Package (HAP), \texttt{Mono} and \texttt{Stereo}. The proposed analysis provides optimal sensitivity over the whole energy range, in contrast to \texttt{Mono} and \texttt{Stereo}, which focus on smaller energy ranges. Improvements in sensitivity of $25-45\%$ are also found for most of the energy range. The analysis is validated using real data from the Crab nebula, showing that the application to data of an IACT analysis can combine significantly different telescope types with significantly different energy ranges. A wider energy coverage, a lower energy threshold, a smaller statistical uncertainty for reconstructed spectral parameters, and a higher robustness are observed. The need for a run-by-run correction for the observation conditions is also highlighted.}

\date{\today}

%\linenumbers

\maketitle

\section{Introduction}
\label{sec:intro}

A few decades after their initial developments~\citep{Weekes1989}, imaging atmospheric Cherenkov telescopes (IACTs) are well established as the primary instrument for studying gamma rays in the energy range of tens of GeV to hundreds of TeV~\citep{Funk_2015}. The current generation is formed by arrays that have been operating for the last several decades, such as the High
Energy Stereoscopic System (H.E.S.S.)~\citep{HessCrab}, the Major Atmospheric Gamma-Ray Imaging Cherenkov
Telescopes (MAGIC)~\citep{MAGIC}, and the Very Energetic Radiation Imaging Telescope Array System (VERITAS)~\citep{weekes2002veritas}. The next generation, led by the Cherenkov Telescope Array Observatory (CTAO)~\citep{CTAConsortium:2013ofs,CTAConsortium:2017dvg}, is expected to become operational during the next years. In the energy range above a few dozen GeV, the usual gamma-ray flux is often too faint to be detected by space-based detectors with effective areas of a few square meters. Ground-based IACTs exploit the air showers that are initiated when the gamma ray interacts with the atmosphere. Charged secondary particles emit Cherenkov light, which is then reflected by large mirrors and collected by the camera in a time window of a few dozen nanoseconds. The spatial and timing structures of the camera image are used to reconstruct the energy and direction of the primary gamma ray and to reject hadron-induced air showers~\citep{HessCrab2006}.

Different telescope sizes employed together can significantly extend the energy range of the experiment. 
This approach will be used by CTAO and has been pioneered by the H.E.S.S. experiment, which currently employs four $12 \, \rm{m}$ telescopes (CT1 to CT4) in a $120 \, \rm{m}$ square together with a central $28 \, \rm{m}$ telescope (CT5). When running in stereoscopic mode (i.e., with the requirement that at least two telescopes participate in an event), H.E.S.S. achieves a competitive performance in the energy range between 150 GeV and 100 TeV~\citep{Krawczynski:2006pd,HessCrab2006,Parsons:2014voa,MAGIC:2014zas,Kosack:2020joq}. With CT5 in monoscopic mode, the energy threshold is lowered to $\sim 30 - 50 \, \rm{GeV}$~\citep{Murach2015}, but the performance is significantly reduced for energies above $\sim 200 \, \rm{GeV}$. For the usual monoscopic and stereoscopic analysis methods, a choice between a low-energy threshold and the overall performance above 200 GeV must therefore be made, depending on the science case.

We propose an analysis based on event types for H.E.S.S. that provides a single combined analysis that can exploit the whole energy range of the experiment with an optimal performance. This approach has been widely used by Fermi-LAT, with a significant performance boost~\citep{Fermi-LAT:2013jgq,Atwood:2013dra}. Nevertheless, considerable adaptations are needed for H.E.S.S. because of the different telescope types and observation multiplicities, and these are explored in this work.

The simulations and datasets we used are presented in section~\ref{sec:dataset}, followed by a description of the definition of the event types in section~\ref{sec:types}. The performances are evaluated and addressed in section~\ref{sec:performance}. The implementation is validated using real data through an analysis of data taken on the Crab nebula, which is shown in section~\ref{sec:crab}. Systematic uncertainties of the observation conditions are discussed in section~\ref{sec:systematics}. Finally, we conclude in section~\ref{sec:conclusions}.

\section{Simulations and dataset}
\label{sec:dataset}

H.E.S.S. is an IACT system located in the Khomas highlands in Namibia at an altitude of $1800 \, \rm{m}$ above sea level. The array has been operating since 2003, initially with four $12 \, \rm{m}$ telescopes, and since 2012, with an added $28 \, \rm{m}$ central telescope~\citep{vanEldik2016}. The monoscopic trigger of CT5 provides a significant reduction in energy threshold~\citep{Murach2015,Unbehaun:2025skr}. In 2019, CT5 was upgraded with the installation of a FlashCam (FC) camera~\citep{FC-Bi-2021}, with upgraded electronics design. This work focuses on the time period after the FC installation.

Gamma-ray-induced air showers and the telescope response to them were simulated using the packages \texttt{CORSIKA}~\citep{Heck:1998vt} and \texttt{sim\_telarray}~\citep{simtel-2008}. A power-law energy spectrum with $dN/dE \propto E^{-1.8}$ was used in the simulations, together with an energy and zenith-dependent importance sampling of simulated impact distances. During the gamma-hadron separation and sensitivity estimation steps, a re-weighting was applied to obtain $E^{-2.5}$ and $E^{-2}$ spectra. Real observational data with masked known gamma-ray sources and bright stars (off runs) were used for the background because hadron-induced air showers constitute the overwhelming background for IACTs. Using off runs in the training minimizes the effects of possible differences between Monte Carlo (MC) simulations and real data~\citep{shilon-2019}. For the validation shown in section~\ref{sec:crab}, we used 27 observation runs (28 min each, and a total of $\sim 12.6$h) of the Crab nebula taken between September 2020 and October 2021 in a zenith angle range from $44^{\circ}$ to $55^{\circ}$. The dataset is a smaller subset of the dataset used in \cite{HessCrab} and in \cite{Unbehaun:2025skr}.

The MC simulations and data were processed with the H.E.S.S. analysis program (\texttt{HAP}). After calibration, the amplitude and peak time of the signal pulse were calculated for each pixel~\citep{HESS:2004wzg,HESS:2021rot}. A cleaning procedure was then applied to remove noise pixels. We used two different cleaning methods, tailcuts and time-based cleaning. The former is the standard cleaning method used in H.E.S.S. and is based on a two-threshold cut. A pixel is considered signal when it exceeds the primary threshold and a given number of its neighboring pixels (usually one or two) exceeds the secondary threshold. In this case, the main pixel and the neighbors that exceed the secondary threshold are kept~\citep{HessCrab2006}. We used thresholds of 5 and 10 photoelectrons (p.e.) with one neighbor, labeled 05/10, for the small telescopes and 09/16NN2 (i.e., with two neighbors) for CT5. We also explored using time-based cleaning as presented in~\cite{Celic:2025wyp} instead of tailcut cleaning for CT5. For this method, the timing correlation of the pixels was also taken into account, which resulted in a multidimensional clustering of the image. An improved performance is expected at the lowest energies due to more signal retention and better gamma-hadron separation. We used the parameters labeled \texttt{TIME3D\_1} in~\cite{Celic:2025wyp}. Finally, for each cleaning, low-level variables such as Hillas parameters~\citep{Hillas} and the variables introduced in~\cite{Unbehaun:2025skr} were calculated and stored.

\section{Event types definition}

Two main analysis configurations are currently available in the \texttt{HAP} analysis pipeline in H.E.S.S.: \texttt{Mono}, in which CT5 is used in monoscopic trigger mode and data measured by CT1-4 are ignored, and \texttt{Stereo}, in which stereoscopic observation of an event (i.e., at least two telescopes participate) is required. The current H.E.S.S. stereoscopic configuration within HAP ignores CT5 data because it is hard to combine different telescope types with different systematics and triggers. \texttt{Mono} can achieve a significantly lower energy threshold because the CT5 collection area is large. Monoscopic reconstruction is very challenging, however, in particular, when the degeneracy between events with lower energy or larger impact distance is to be distinguished, which leads to a deterioration of the energy and angular resolutions. The effective area for \texttt{Mono} is much reduced compared to that of \texttt{Stereo} at $E \gtrsim 200~\rm{GeV}$. This is expected because a larger ground area is covered when the four small telescopes are used. For \texttt{Stereo}, the overall performance for energies above a few hundred GeV is naturally improved, but the lowest energies become inaccessible. Combining \texttt{Mono} and \texttt{Stereo} analyses in a single joint-likelihood fitting approach to obtain an optimal performance over the whole energy range is not straightforward because events can belong to \texttt{Mono} and \texttt{Stereo}. For example, many events are detected by CT5 and at least two small telescopes, which are counted by both \texttt{Mono} and \texttt{Stereo} analyses. An approach based on choosing the best reconstruction for overlapping events was explored in~\cite{vanEldik2016}, but no sensitivity improvements for to \texttt{Mono} and \texttt{Stereo} were found.

We propose here a method capable of exploiting the whole energy range of H.E.S.S. with an optimal performance in a single analysis by using event types. The method is based on the concept of the event types used by Fermi-LAT, which allows for an optimized reconstruction for different categories of events~\citep{Fermi-LAT:2013jgq}. In Fermi-LAT, these categories are based on the quality estimation of a first reconstruction. For H.E.S.S., however, different telescopes are used, and some events can therefore be reconstructed stereoscopically, while others must be reconstructed monoscopically. For this reason, it is not straightforward to define the types with reconstructed quantities. This has been explored in the past~\citep{vanEldik2016,HESS:2019rhe}, but a different approach was chosen for this analysis. 

We introduce a definition of types that is based on the Hillas parameters, in particular, the size, that is, summed amplitude of the pixels surviving cleaning, the number of pixels after cleaning, and the local distance, that is, the distance of the center of gravity of the cleaned image to the center of the camera. Table~\ref{tab:cuts} shows the preselection cuts that define each event type. The labeling of the types is sequential. An event can only be labeled B if it was not labeled A, and so on. This ensured nonoverlapping types.

\label{sec:types}
\begin{table*}[ht!]
    \centering
        \caption{Preselection cuts that define each event type.}
    \label{tab:cuts}

\begin{tabular}{c|c|c|c|c|c|c|c|c|c|c|c}
\hline
\hline
\multirow{4}{*}{Type} & \multicolumn{8}{c|}{Preselection cuts on Hillas image} & \multicolumn{3}{c}{Requirements} \\
\cline{2-12}
& \multicolumn{2}{c|}{Cleaning} & \multicolumn{2}{c|}{Size [p.e.]} & \multicolumn{2}{c|}{\# pixels} & \multicolumn{2}{c|}{Local distance [m]} & & \multicolumn{2}{c}{\makecell{Multiplicity\\(after preselection)}} \\ % <-- changed here
\cline{2-9}
\cline{11-12}
& CT1-4 & CT5 & CT1-4 & CT5 & CT1-4 & CT5 & CT1-4 & CT5 & & CT1-4 & CT5 \\
\hline
\hline
\texttt{A} & Tailcuts & Tailcuts & $\ge 200$ & $\ge 50$ & $>0$ & $> 0$ & $\le 0.525$ & - & & $\ge 2$ & $=1$ \\
\texttt{B} & Tailcuts & Tailcuts & $\ge 80$ & $\ge 80$ & $>0$ & $> 0$ & $\le 0.525$ & $\le 0.8$ & not \texttt{Type A}, and & \multicolumn{2}{c}{$\ge 2$} \\
\texttt{M} & Tailcuts & Time-based & $\ge 80$ & $\ge250$ & $>0$ & $\ge 7$ & $\le 0.525$ & $\le 0.8$ & not \texttt{Type A}, \texttt{B}, and & $=0$ & $=1$ \\
\hline
\hline
\end{tabular}

\tablefoot{For the tailcut cleaning, we used the standard H.E.S.S. thresholds of 5/10 for CT1-4 and 9/16 with a two-neighbor requirement for CT5. For the time-based cleaning, we used \texttt{TIME3D\_1} for \texttt{Type M} (see~\cite{Celic:2025wyp} for a detailed description of the time-based cleaning parameters).
}
    
\end{table*}

\begin{figure}
    \centering
    \includegraphics[width=0.98\linewidth]{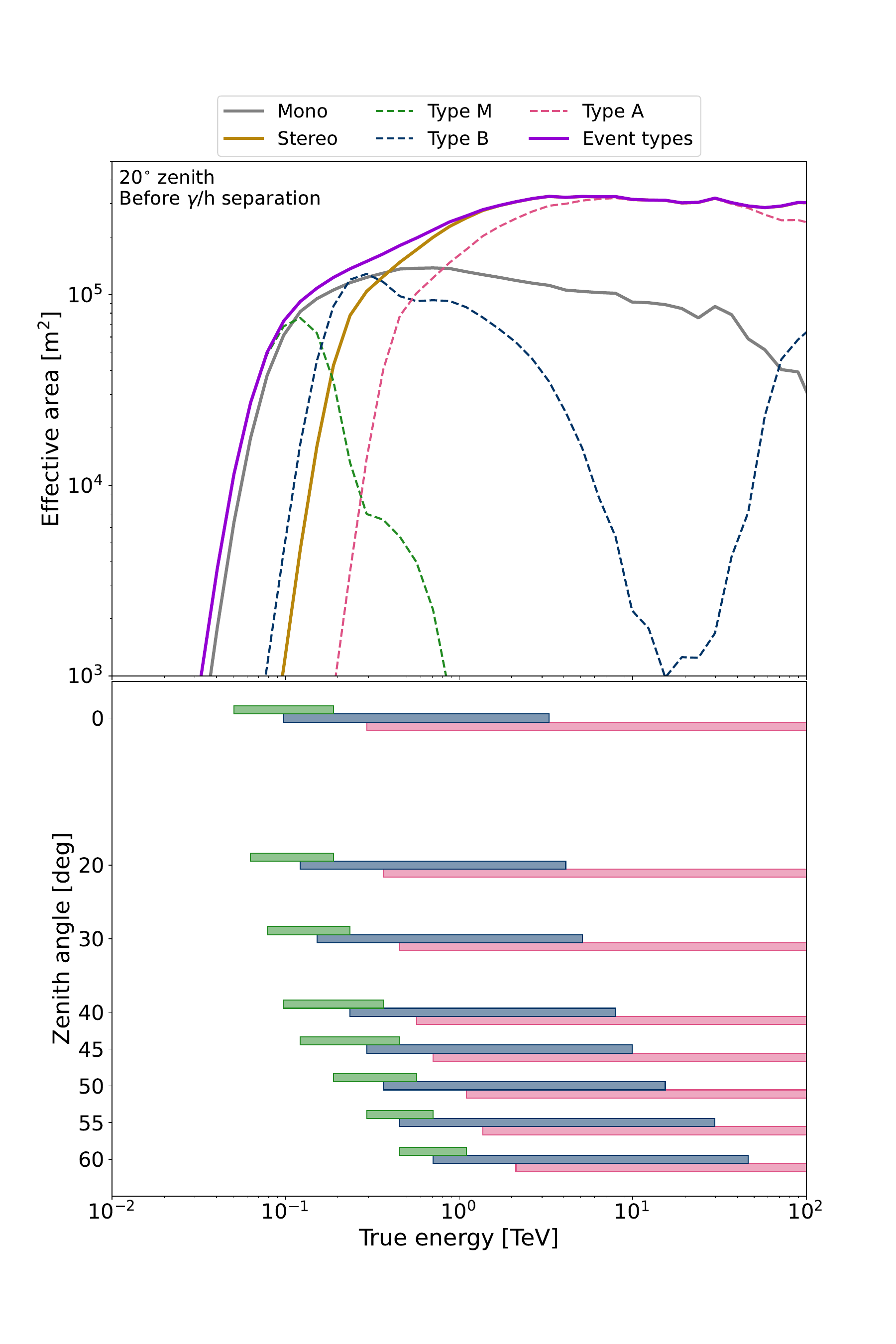}
    \caption{Top panel: Effective area before gamma-hadron separation for a representative zenith angle of $20^{\circ}$. The standard configurations of \texttt{HAP}, \texttt{Mono} (full gray line) and \texttt{Stereo} (full orange line) are compared to the types \texttt{Type M} (dashed green line), \texttt{Type B} (dashed blue line), and \texttt{Type A} (dashed magenta line) and to the combined \texttt{Event types} (full purple line). Bottom panel: Energy ranges within which each type contributes to at least 10\% of the total effective area for different zenith angles. The lower limit of \texttt{Type M} is defined as the energy for which its effective area drops below 10\% of the maximum of the combined one.}
    \label{fig:typefraction}
\end{figure}

Figure~\ref{fig:typefraction} compares the energy range coverage between the standard configurations and event types. Effective areas were obtained considering only preselection cuts, and without gamma-hadron separation and energy and angular reconstructions. The combination of event types can provide a full coverage of the energy range, in contrast to the standard configurations, \texttt{Mono} and \texttt{Stereo}. \texttt{Type A} contains the best stereoscopic events, requiring CT5 and two images on CT1-4 with more than 200 p.e. and dominating at the highest energies ($E \gtrsim 500 \, \rm{GeV}$). \texttt{Type B} covers the lower-quality stereoscopic events, peaking in the energy range between $\sim 150 \, \rm{GeV}$ and $\sim 500 \, \rm{GeV}$. The majority of the high-energy events are detected by CT5, and are thus part of \texttt{Type A}. Nevertheless, a further recovery of the \texttt{Type B} distribution is seen beyond 30 TeV because very energetic events fall far from the center of the array. These events are bright enough to be detected by two outer telescopes, but are far enough away not to be detected by CT5. Finally, \texttt{Type M} takes into account the true monoscopic events, that is, events detected by CT5 that are not detected by any of the small telescopes, dominating at energies below $\lesssim 150 \, \rm{GeV}$. The presented energy ranges are valid for a representative zenith angle of $20^{\circ}$, as is common for estimating and comparing IACT performances. As shown in the bottom panel of Figure~\ref{fig:typefraction}, larger zenith angles lead to a shift of the threshold toward higher energies.

In addition to the definition of the types, we introduce several improvements to the reconstructions and gamma-hadron separation, as outlined below. To isolate the impact of the event types from that of these enhancements, we also evaluate the performance of analysis configurations that incorporate these improvements, but are based on the old definitions of \texttt{Mono} and \texttt{Stereo}. These are referred to as \texttt{Mono++} and \texttt{Stereo++}.

\subsection{Improvements to monoscopic configurations (\texttt{Type M} and \texttt{Mono++})}
\label{sec:mono}

The monoscopic improvements are based on~\cite{Unbehaun:2025skr} and ~\cite{Celic:2025wyp} and include the following four steps:
\begin{itemize}
    \item Variables that complement the traditional Hillas parameters with increased sensitivity to the time and intensity distributions in the image are added. These variables are used in the angular and energy reconstruction and in the gamma-hadron separation. An improvement of 25-40\% in retention of gamma-ray signal for a similar background rejection is seen~\citep{Unbehaun:2025skr}.
    \item A neural network (NN) for determining the correct orientation of the image in the camera frame is added. For monoscopic reconstructions, it is not enough to determine the major axis of the image to break the degeneracy between the head and the tail of the ellipse. Improvements of 57\% in angular resolution are seen at energies $\sim 100 \, \rm{GeV}$~\citep{Unbehaun:2025skr}.
    \item A size-dependent cut for the gamma-hadron separation is added. This ensures optimal performance in different regimes and leads to a reduction of $\sim 50\%$ in the energy threshold~\citep{Unbehaun:2025skr}.
    \item The use of time-based cleaning instead of tailcut cleaning. The information about the time distribution can substantially contribute to retaining more signal while better rejecting background pixels for faint images in particular. A $\sim 15\%$ improvement is seen in the sensitivity below 300 GeV, combined with a significant energy threshold reduction~\citep{Celic:2025wyp}.
\end{itemize}

\subsection{Improvements to stereoscopic configurations (\texttt{Type A}, \texttt{Type B}, and \texttt{Stereo++})}
\label{sec:hybrid}

Three main improvements are introduced for the stereoscopic configurations. They are described below.

\begin{itemize}
    \item The addition of CT5, which combines the two telescope types. This was possible through updated MC simulations for CT5 after an extensive data-MC validation procedure~\citep{Leuschner:2023ega}, a fine-tuning of the size threshold of each telescope to avoid systematics arising from different triggers, and separation of the mean-scaled variables (as discussed below).
    \item The separation of the usual mean-scaled variables~\citep{Daum:1997vf,HessCrab2006} into mean-scaled variables considering CT1-4 alone and individual Hillas variables for CT5. While the mean-scaled variables are designed to follow a normal distribution for signal events, their distribution for background events would strongly depend on the multiplicity when different telescope types are combined, leading to performance losses. The list of final variables used in the boosted decision tree (BDT) is presented in Appendix~\ref{app:BDT}.
    \item A reconstructed energy-dependent cut for the gamma-hadron separation. For each bin of reconstructed energy, the BDT cut that optimizes the q-factor (gamma efficiency over the square root of background survival rate) is chosen. The cut evolution with energy is then smoothed. This is similar to the procedure introduced in~\cite{Unbehaun:2025skr}. Because multiple telescopes are involved, reconstructed energy was used instead of Hillas size, however.
\end{itemize}

\section{Performance}
\label{sec:performance}

We present the performances for the current standard \texttt{HAP} configurations (\texttt{Mono} and \texttt{Stereo}), the configurations that follow the old definition, but use the introduced improvements (\texttt{Mono++} and \texttt{Stereo++}), and the types that follow the new definitions and improvements we propose here (\texttt{Type A}, \texttt{Type B}, and \texttt{Type M}, and when applicable, the combined \texttt{Event types}). A representative zenith angle of $20^{\circ}$ is shown.

\subsection{Effective area}

\begin{figure}
    \centering
    \includegraphics[width=0.98\linewidth]{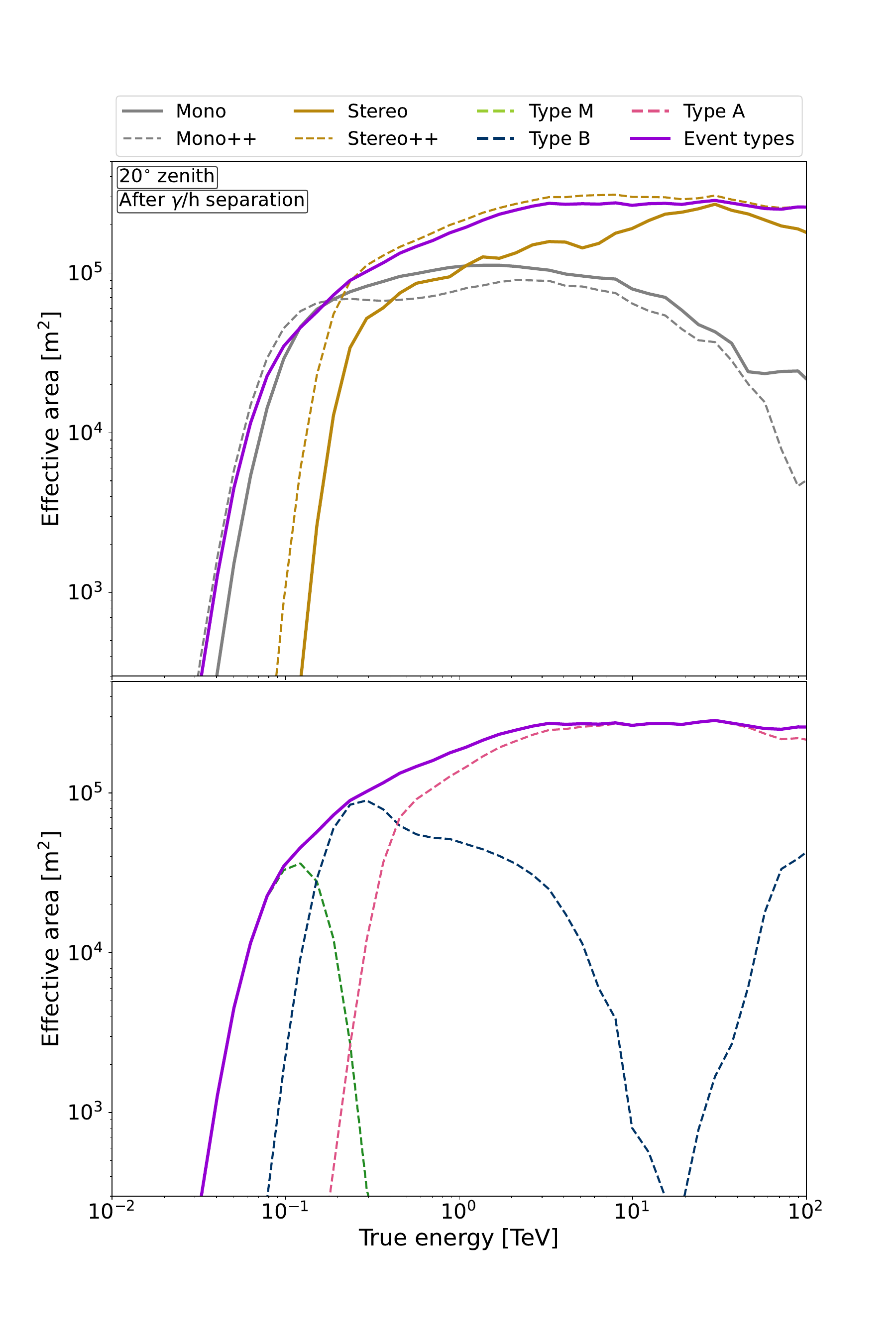}
    \caption{Effective area after gamma-hadron separation for a representative angle of $20^{\circ}$. In the top panel, \texttt{Event types} (full purple line) is compared to the standard configurations of \texttt{HAP}, \texttt{Mono} (full gray line), and \texttt{Stereo} (full orange line), and to the intermediate configurations with the old definition and new improvements, \texttt{Mono++} (dashed gray line) and \texttt{Stereo++} (dashed orange line). In the bottom panel, \texttt{Event types} is compared to the individual types \texttt{Type M} (dashed green line), \texttt{Type B} (dashed blue line), and \texttt{Type A} (dashed magenta line).}
    \label{fig:aeff}
\end{figure}

The effective area after gamma-hadron separation is shown in Figure~\ref{fig:aeff}. A significantly lower energy threshold is seen in \texttt{Mono++} with relation to \texttt{Mono}. This is mostly due to the size-dependent BDT cut, which leads to a higher gamma efficiency, especially at the lowest energies. \texttt{Stereo++} also shows a lower energy threshold and slightly larger effective area up to $\sim 10 \, \rm{TeV}$. The improvement in energy threshold is due to the addition of CT5, which is able to detect fainter showers, while the improvement at the intermediate energies is due to the improved gamma-hadron separation. 
Still, neither one of the improved \texttt{Mono++} and \texttt{Stereo++} can provide an optimal performance over the whole energy. This can only be achieved with \texttt{Event types}, with each type dominating different energy ranges, as shown in the bottom panel.
A small reduction in effective area is seen in \texttt{Event types} with respect to \texttt{Mono++} and \texttt{Stereo++} due to harder gamma-hadron separation cuts arising from the cut optimization, as shown in Section~\ref{sec:eff}. This does not lead to performance losses because the cuts optimize signal over noise rather than signal retention.

\subsection{Defining a combined resolution for \texttt{Event types}}
\label{sec:combined}

\begin{figure}
    \centering
    \includegraphics[width=0.98\linewidth]{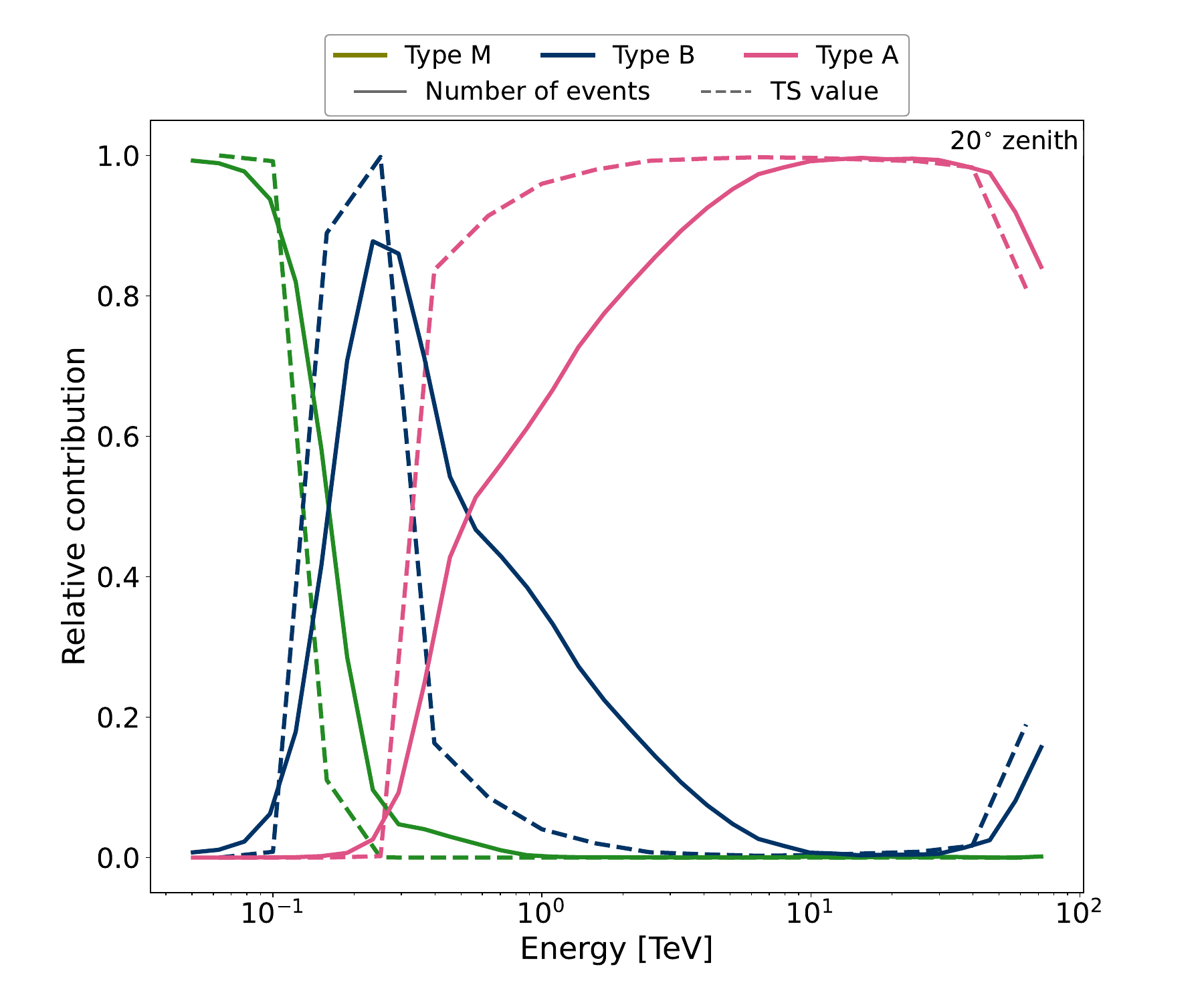}
    \caption{Relative contribution of each type to the total number of events (full lines) and the test statistic value for a gamma-ray source with flux at the sensitivity level (dashed lines) for a representative angle of $20^{\circ}$. \texttt{Type M}, \texttt{Type B}, and \texttt{Type A} are represented by green, blue, and magenta lines, respectively.}
    \label{fig:ts}
\end{figure}

While the combined effective area for \texttt{Event types} can easily be obtained by simply summing the effective areas from \texttt{Type M}, \texttt{Type B}, and \texttt{Type A}, a similar approach is not possible for the combined separation efficiencies and reconstruction resolutions. Event-types-based analyses are designed such that the reconstruction for events in each type is performed independently and followed by a joint likelihood fit. Due to the asymmetry of high- and low-quality events in the types, a given type can contribute considerably more to the statistical significance of the fit, even with a similar contribution to the number of events. This is illustrated in Figure~\ref{fig:ts}, in which the relative contribution of each type to the total number of events is compared to the relative contribution to the test statistic (TS) value for a gamma-ray source with flux on the sensitivity level. The statistical assumptions for the joint likelihood fit follow those described in Section~\ref{sec:sens}, and TS was defined as $TS = -2 \log{\mathcal{L}_0} - (-2 \log \mathcal{L}_1)$, with $\mathcal{L}_0$ as the null hypothesis and $\mathcal{L}_1$ as the optimized likelihood. About $500 \, \rm{GeV}$, \texttt{Type B} and \texttt{Type A} contribute equally to the effective area. Because of its higher-quality events, however, \texttt{Type A} contributes about nine times more to the TS value. The transition between the dominance of \texttt{Type M} to \texttt{Type B} on the relative contribution to TS value is seen at $E \sim 140 \, \rm{GeV}$, while the transition between \texttt{Type B} and \texttt{Type A} occurs at $E \sim 350 \, \rm{GeV}$. When the effective combined efficiencies and resolutions shown in the following sections are determined, we combined all the events in a single dataset and weighted them by the relative TS contribution of that type for the given energy bin. True energy was used for the resolutions, and reconstructed energy was used for the q-factor. This is justifiable given the well-behaved energy resolution and bias. The uncertainties in the efficiencies and resolutions were calculated via bootstrapping by resampling 20\% of the data 100 times and calculating the mean and standard deviation. For \texttt{Event types}, sharp changes are seen because TS changes rapidly in the energy ranges within which the transitions between the types occur.

The reasoning for the re-weighting of resolutions with TS was to obtain an effective resolution as would be perceived by a joint fit. In the joint fit, the TS values of the individual types are added, and the combined TS is minimized. The effect of each type on the combined fit is therefore proportional to its TS, that is, a poor resolution for a type with relatively small TS will not deteriorate the final fit, and the type with the highest TS value will dominate the resolution. For this reason, the resolutions as presented here are representative of resolutions that can be extracted from the likelihood surface. They are therefore the resolutions that better estimate how well the analysis will perform because they are those that should be compared to those of the standard methods.

\subsection{gamma-hadron separation efficiencies}
\label{sec:eff}

\begin{figure*}
    \centering
    \includegraphics[width=0.92\linewidth]{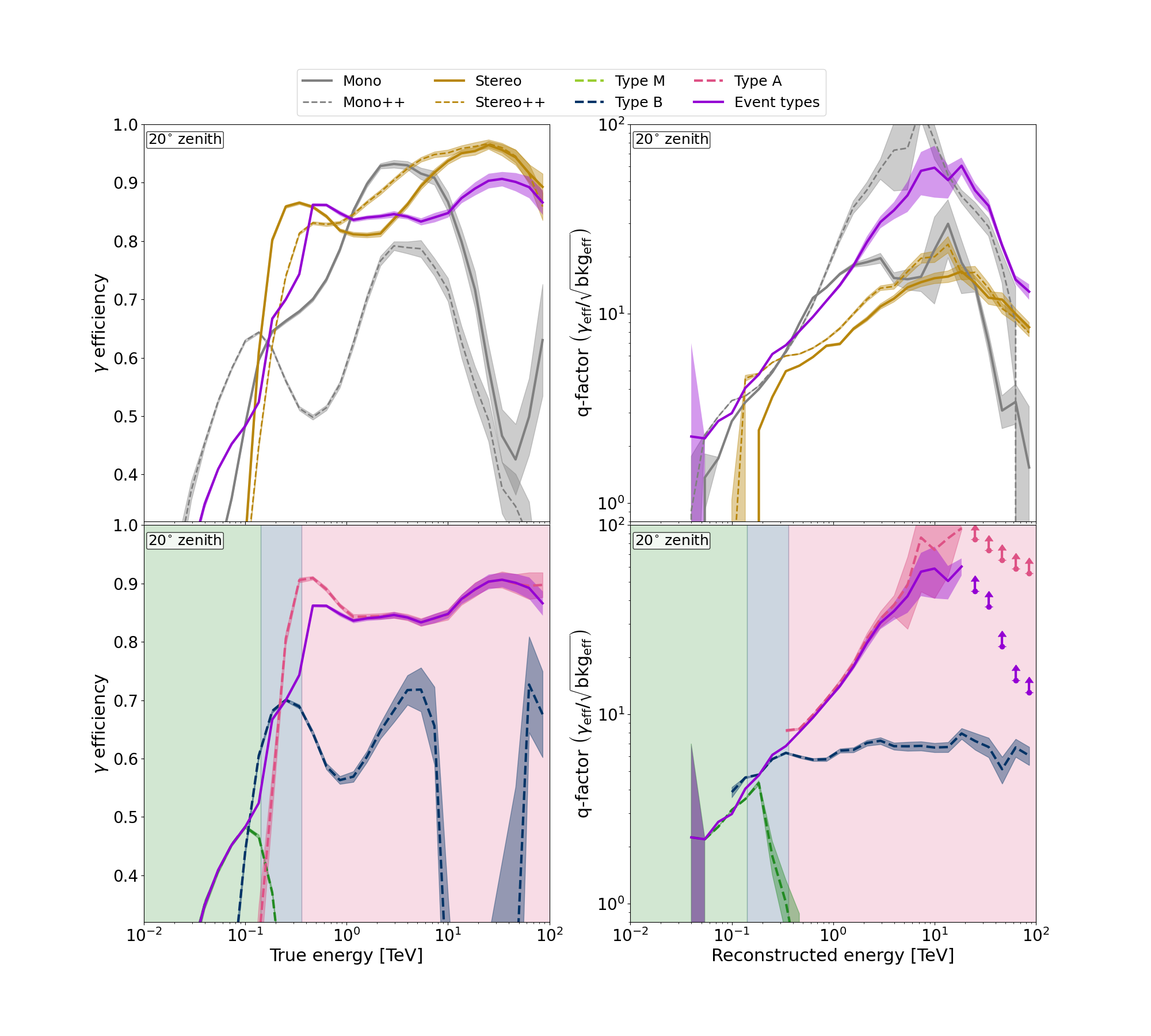}
    \caption{Separation efficiencies as a function of energy for a representative zenith angle of $20^{\circ}$. The left panels show the gamma efficiency, $\gamma_{\rm{eff}}$, as a function of true energy. The right panels show the q-factor, $\gamma_{\rm{eff}}/\sqrt{\rm{bkg}_{\rm{eff}}}$, as a function of reconstructed energy (as real data were used for the background, true energy is not defined). In the top panels, \texttt{Event types} (full purple line) is compared to the standard configurations of \texttt{HAP}, \texttt{Mono} (full gray line), and \texttt{Stereo} (full orange line) and to the intermediate configurations with the old definition and new improvements, \texttt{Mono++} (dashed gray line) and \texttt{Stereo++} (dashed orange line). In the bottom panels, \texttt{Event types} is compared to the individual types: \texttt{Type M} (full green line), \texttt{Type B} (full blue line), and \texttt{Type A} (full magenta line). The uncertainties are calculated via bootstrapping and are shown as shaded regions around the lines. The shaded areas in the bottom panel show the energy range in which each type dominates the joint analysis. The combined efficiency for \texttt{Event types} was calculated as discussed in Section~\ref{sec:combined}.}
    \label{fig:eff}
\end{figure*}

Figure~\ref{fig:eff} shows the gamma-hadron separation performance. The gamma efficiency, $\gamma_{\rm{eff}}$, is defined as the ratio of events left after BDT cuts. Softer cuts lead to higher gamma efficiencies, that is, to a higher signal retention, but also to higher background survival rates, that is, to background contamination. The opposite is seen for harder cuts. To first order, an optimal cut maximizes the q-factor, $\gamma_{\rm{eff}}/\sqrt{\rm{bkg}_{\rm{eff}}}$. When compared to the standard monoscopic configuration, \texttt{Mono}, the improved types, \texttt{Mono++} and \texttt{Event types}, retain more signal and background at the lowest energies, but still result in a higher q-factor. This is due to the size-dependent BDT cut. In the standard case, a single BDT cut is applied to the whole energy range. This cut value is optimized such that the integral sensitivity is maximized, resulting in suboptimal performance at the lowest and highest energies. In contrast, a varying BDT cut can ensure an optimal performance over the whole energy range, and to do this, a soft cut is needed for low-size images, for which the separation is more challenging. 

In the stereoscopic case, \texttt{Stereo++} shows a slightly improved q-factor with relation to \texttt{Stereo} in almost the whole energy range and reaches lower energies. This comes from a combination of using CT5, improved BDT variables, and an energy-dependent BDT cut. \texttt{Event types} performs significantly better than \texttt{Stereo++} for most of the energy range. It achieves q-factors that are two to three times better for a few TeV. This is due to the dominance of \texttt{Type A}, for which the high-quality events are much better separated. \texttt{Type B} events, even though they contribute to 10-30\% of the total events in this energy range, are neglected by a combined analysis and therefore do not lead to a deterioration of the performance as for \texttt{Stereo++}. A strong dip in the gamma efficiency for \texttt{Type B} is seen around $10 \, \rm{TeV}$, resulting in values below the required $60\%$. This is due to the poor energy resolution for \texttt{Type B} at these energies (see Fig.~\ref{fig:enres}). While the gamma efficiencies in Fig.~\ref{fig:eff} are presented as a function of true energy, the cuts are defined as a function of the reconstructed energy. Nevertheless, the effect of this dip is minimal given the negligible effective area of \texttt{Type B} in this energy range (as shown in Figure~\ref{fig:aeff}).

For $E \gtrsim 400$ GeV, the q-factor for \texttt{Mono++} is higher than that for the stereo configurations, which indicates that the monoscopic separation might outperform the stereoscopic separation. Nevertheless, as shown in Fig.~\ref{fig:aeff}, the selection of events for \texttt{Mono++} is vastly stricter for these energies, and only high-quality events are kept, which leads to an artificially higher q-factor.

\subsection{Reconstruction resolutions}

\begin{figure}
    \centering
    \includegraphics[width=0.98\linewidth]{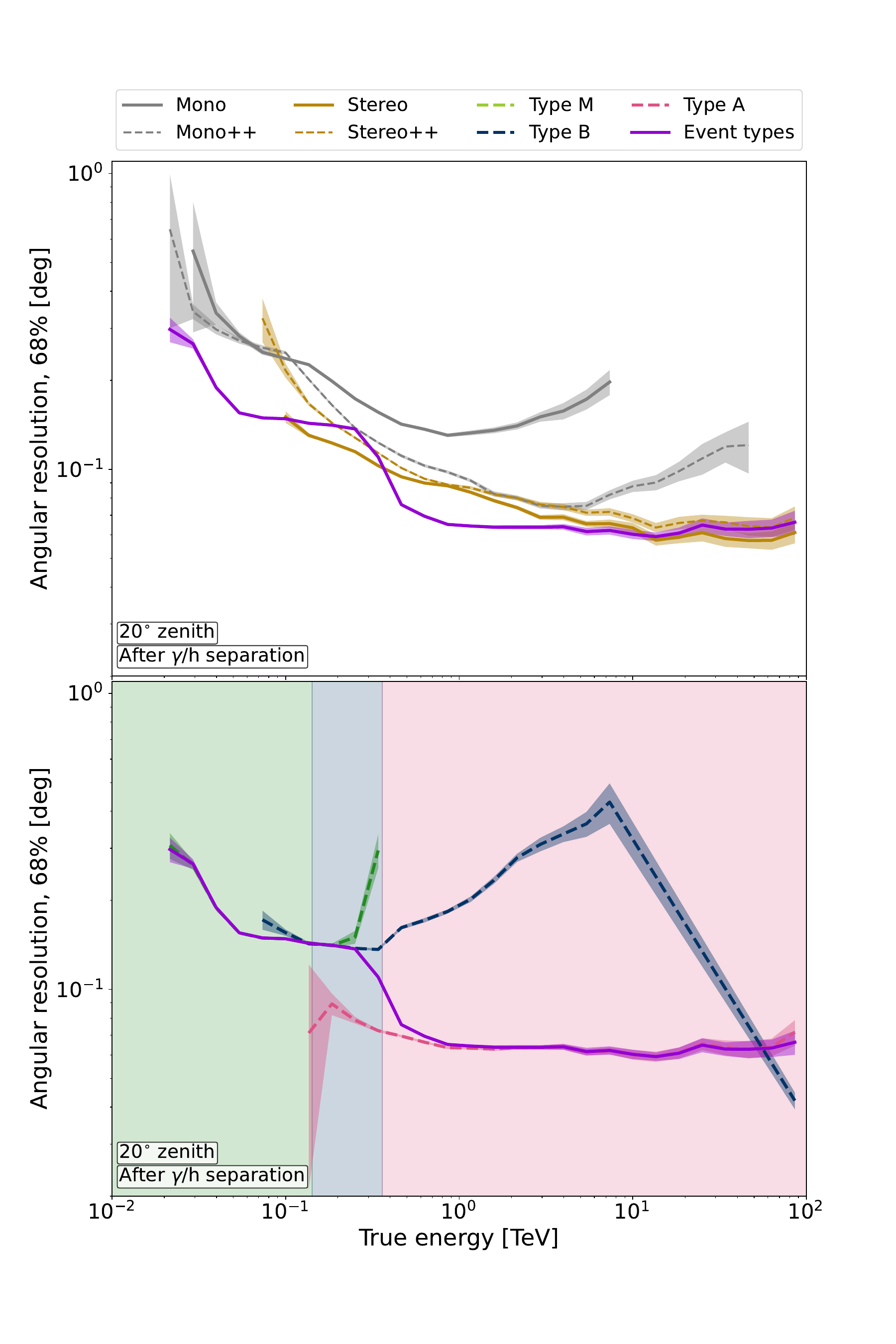}
    \caption{Angular resolution, defined as the 68\% containment radius, as a function of true energy for a representative zenith angle of $20^{\circ}$. In the top panel, \texttt{Event types} (full purple line) is compared to the standard configurations of \texttt{HAP}, \texttt{Mono} (full gray line), and \texttt{Stereo} (full orange line) and to the intermediate configurations with the old definition and new improvements \texttt{Mono++} (dashed gray line), and \texttt{Stereo++} (dashed orange line). In the bottom panel, \texttt{Event types} is compared to the individual types \texttt{Type M} (full green line), \texttt{Type B} (full blue line), and \texttt{Type A} (full magenta line). The uncertainties are calculated via bootstrapping and shown as shaded regions around the lines. The shaded areas in the bottom panel show the energy range in which each type dominates the joint analysis. The combined resolution for \texttt{Event types} was calculated as discussed in Section~\ref{sec:combined}.}
    \label{fig:angres}
\end{figure}

\begin{figure*}
    \centering
    \includegraphics[width=0.92\linewidth]{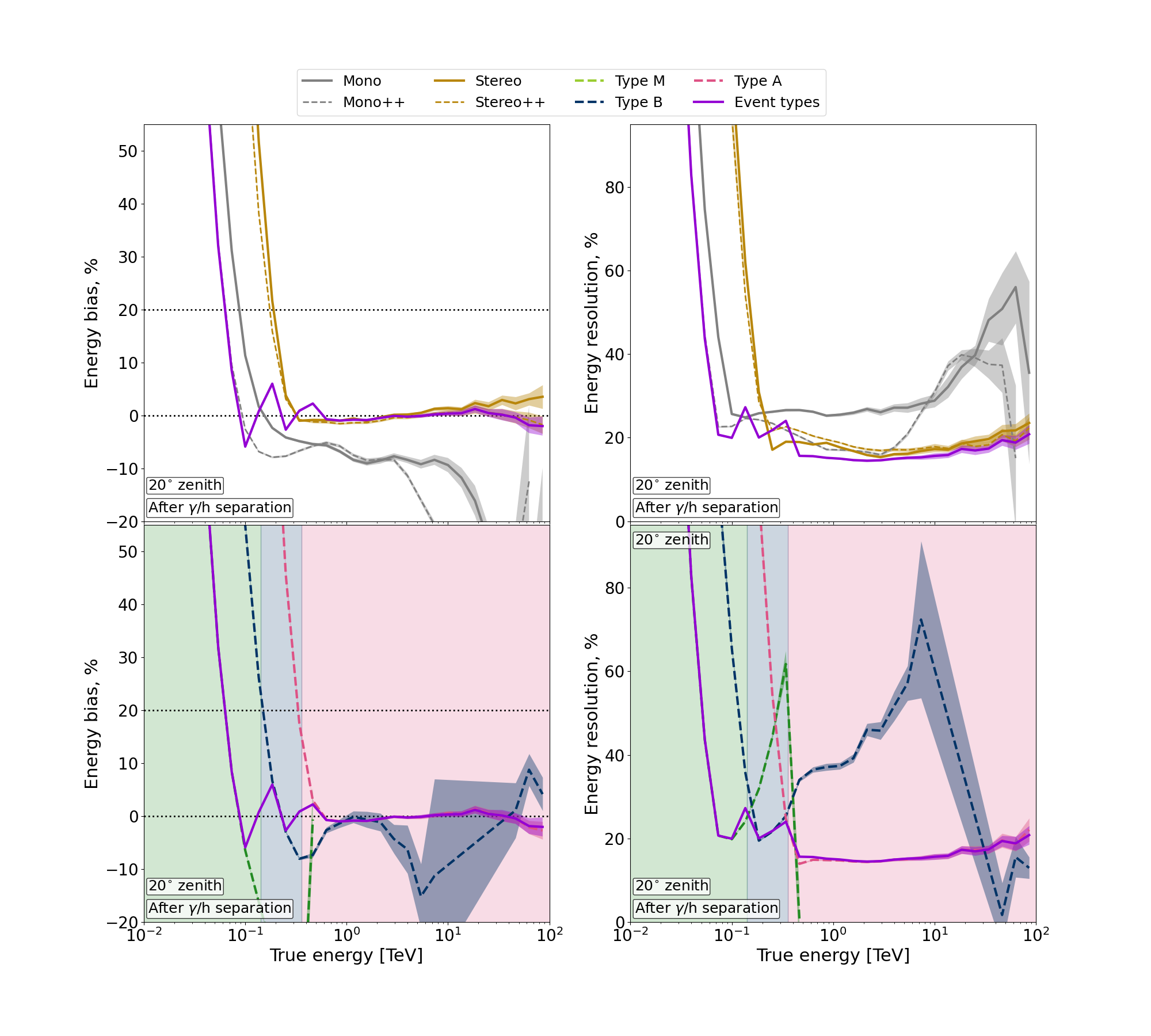}
    \caption{Energy bias and resolution as a function of true energy for a representative zenith angle of $20^{\circ}$. Both are taken from the $(E_{\rm{rec}}-E_{\rm{MC}})/E_{\rm{MC}}$ distribution. Bias (left column) is defined as the median of the distribution, while resolution (right column) is defined as the half-width of the interval around zero that contains 68\% of the distribution. In the top panel, \texttt{Event types} (full purple line) is compared to the standard configurations of \texttt{HAP}, \texttt{Mono} (full gray line), and \texttt{Stereo} (full orange line) and to the intermediate configurations with the old definition and new improvements, \texttt{Mono++} (dashed gray line), and \texttt{Stereo++} (dashed orange line). In the bottom panel, \texttt{Event types} is compared to the individual types \texttt{Type M} (full green line), \texttt{Type B} (full blue line), and \texttt{Type A} (full magenta line). The uncertainties are calculated via bootstrapping and shown as shaded regions around the lines. The shaded areas in the bottom panel show the energy range in which each type dominates the joint analysis. The combined resolution for \texttt{Event types} was calculated as discussed in Section~\ref{sec:combined}.}
    \label{fig:enres}
\end{figure*}

Figures~\ref{fig:angres} and \ref{fig:enres} show the angular and energy resolutions, respectively. The main effect is seen in the monoscopic configurations, as presented in~\cite{Unbehaun:2025skr}. The angular resolution is mostly improved by the NN to reconstruct the image orientation. In monoscopic mode, the primary particle direction is reconstructed by using three estimates in the camera reference frame: the main axis of the ellipse, the distance of the center of the ellipse to the position of the primary particle (which mostly depends on the impact parameter), and the relative orientation between the ellipse center and the primary direction (\texttt{flip}). Since the last one is binary, even with a good estimate of the others, a poor \texttt{flip} estimation can lead to a significantly large angular distance between the reconstructed and true direction, which degrades the angular resolution. The NN reduces the fraction of false \texttt{flip} events by 25-80\%, depending on the energy. The energy resolution is improved by introducing time-sensitive variables, which help to break the degeneracy on the Hillas size between energy and impact parameter. For stereoscopic configurations, angular and energy reconstructions are performed geometrically and through lookup tables~\citep{HessCrab2006}, respectively. This is not changed, and no significant change is therefore observed in individual events. For energies of about $300 \, \rm{GeV}$, \texttt{Event types} and \texttt{Stereo++} present slightly worse resolutions. This is due to additional faint events that did not survive the cuts using only the small telescopes. For energies above this, most poorly reconstructed events belong to \texttt{Type B}, which is neglected by the combined analysis at these energies. Small improvements are therefore seen in both angular and energy resolutions, especially between $\sim 500 \, \rm{GeV}$ and $\sim 3 \, \rm{TeV}$. Even further improvements in energy and angular reconstructions are expected if template-based methods, such as the Image Pixel-wise fit for Atmospheric Cherenkov Telescopes (ImPACT; \citealt{Parsons:2014voa}), are used. This method is commonly used in H.E.S.S. analyses and is expected to be incorporated into analyses using event types in the future.

\subsection{Sensitivity}
\label{sec:sens}

\begin{figure}
    \centering
    \includegraphics[width=0.98\linewidth]{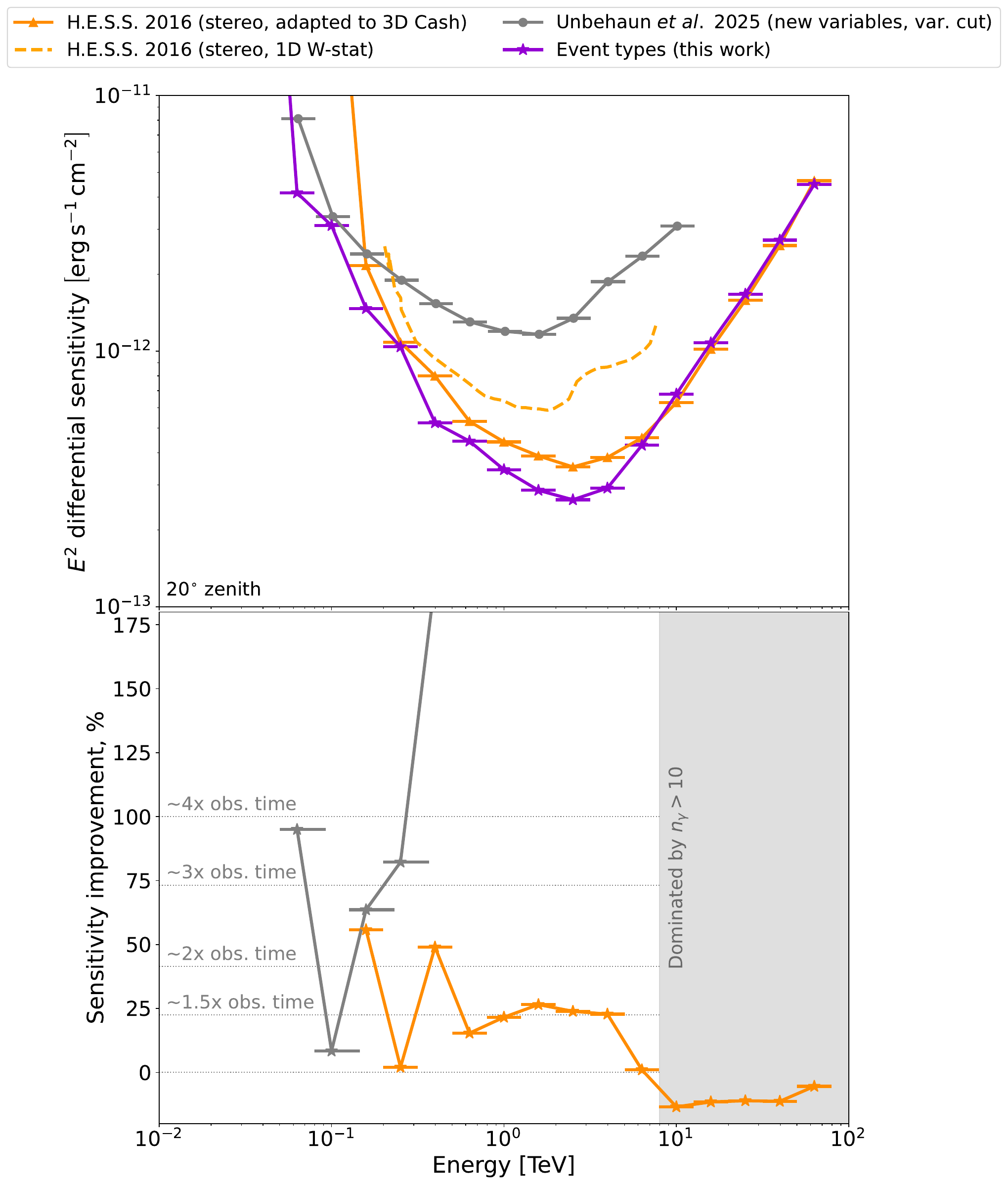}
    \caption{Differential sensitivity as a function of energy for a representative zenith angle of $20^{\circ}$. An energy spectrum of $E^{-2}$, 50h of observation time, a detection significance of 5$\sigma$, more than ten gamma events, and five energy bins per decade are considered. The top panel shows the absolute value for the reference monoscopic (full gray line with circles) and stereoscopic (full orange line with triangles) configuration of \cite{Unbehaun:2025skr} and \cite{vanEldik2016} (both with 1D W-statistics and adapted to 3D Cash statistics) together with the combined \texttt{Event types} (full purple line with stars). The bottom panel shows the relative improvement of \texttt{Event types} with relation to the reference sensitivities. The increase in observation time that would lead to an equivalent sensitivity gain is shown by horizontal dotted lines up to $8 \, \rm{TeV}$. The region above this energy (noted with a gray zone) is dominated by the requirement of more than ten gamma events.}    
    \label{fig:sensitivity}
\end{figure}

A trade-off is observed in the improvements: With an improved gamma-hadron separation, a higher signal retention can be achieved with a similar background rejection. Nevertheless, even with improved reconstruction, a higher signal retention implies a slightly lower resolution because lower-quality events are kept. The combination of these effects in the overall performance can be seen in the differential sensitivity, which estimates the lowest flux that can be detected in an energy bin with the given statistical requirements. Figure~\ref{fig:sensitivity} shows the final sensitivity for \texttt{Event types} compared to the monoscopic and stereoscopic sensitivities from \cite{Unbehaun:2025skr} and \cite{vanEldik2016}\footnote{The data points from Figure~\ref{fig:sensitivity} are listed in Table~\ref{tab:data} in Appendix~\ref{app:data}.}. A point-source detection with a 5$\sigma$ confidence level and more than ten gamma events for an energy spectrum of $E^{-2}$ and 50h of observation time was required. 3D Cash statistics~\citep{Cash} was used instead of the usual 1D W-statistics with on-off regions~\citep{Berge2007}. The reference stereoscopic sensitivity from~\citep{vanEldik2016} was adapted to 3D Cash statistics for Figure~\ref{fig:sensitivity}. The sensitivity for \texttt{Event types} using W-statistics is shown in Appendix~\ref{app:sens}. The differential sensitivity strongly depends on the bin size and the chosen requirements and statistical assumptions. A comparison between the different configurations is therefore valid, but caution is required for a comparison to the sensitivities obtained in other works. For \texttt{Event types}, the datasets for each type were built independently, and then a joint fit was performed to combine them. An observation offset, that is, the angular distance between the expected source position and the pointing direction, of $0.5^{\circ}$ was considered. The variability of the improvements over the field of view of the telescope is shown in Appendix~\ref{app:fov}.

The broad potential of \texttt{Event types} is clear. Optimal performance is found throughout the entire energy range, in contrast to monoscopic and stereoscopic configurations, which are only optimal at $E<200 \, \rm{GeV}$ and $E>200 \, \rm{GeV}$, respectively. The further improvements described in sections~\ref{sec:mono}~and~\ref{sec:hybrid} lead to significant performance improvements, especially in the lowest energies, with the energy threshold reduced from $\sim 80$ to $\sim 50 \, \rm{GeV}$. Moreover, further improvements are seen especially at $0.25 < E/\rm{TeV} < 10$, which arise from the asymmetry in event quality between \texttt{Type B} and \texttt{Type A}, which makes it possible for the fit to neglect low-quality events. This enhances the contribution of high-quality events. This highlights the importance of using event types in addition to the improvements introduced in Sections~\ref{sec:mono} and \ref{sec:hybrid}.

In first order, the sensitivity is expected to improve (i.e., decrease) as $1/\sqrt{t_\mathrm{obs}}$ for energies up to $\sim 8 \, \rm{TeV}$. With this, the equivalent additional observation time needed to achieve improvements on the same order can be estimated. This is also shown in Figure~\ref{fig:sensitivity}. Improvements of $\sim 25-45\%$ are found for most of the energy range, which is equivalent to observing $1.5 - 2$ times as long. For energies above $8 \, \rm{TeV}$, the sensitivity as defined is dominated by the requirement for more than ten gamma events. This is led by the total effective area, which is slightly smaller for \texttt{Event types} than for the usual stereoscopic configurations. Nevertheless, it must be noted that this requirement focuses on a point-source detection in a single energy bin. For usual analyses, the source will likely be observed over many bins. For these cases, the requirement is dropped, and improvements for \texttt{Event types} are seen at these energies as well.

\section{Validation with real data}
\label{sec:crab}

We validated the new configurations using observations of the Crab nebula. The \texttt{GAMMAPY} package~\citep{gammapy:2023} was used for the spectral analysis. Reflected regions~\citep{Berge2007} were used for estimating the background in this validation. Alternative background estimations, such as the 3D background model~\citep{Mohrmann:2019hfq}, can be performed in future analyses. This choice has no significant effect on the validation. On-regions with $0.1^{\circ}$ ($0.2^{\circ}$) radii were used for the stereoscopic (monoscopic) configurations. For \texttt{Mono} and \texttt{Stereo}, a low-energy threshold was set at the energy for which the effective area drops to $10\%$ of its maximum. For \texttt{Type M}, 5\% of the total effective area of all types was used because the energy range covered by \texttt{Event types} is vast. In addition, only the energy ranges for which each type contributed more than $20\%$ of the combined effective area were used in each individual analysis. This resulted in an upper threshold for \texttt{Type M}, an upper and a lower threshold for \texttt{Type B}, and a lower threshold for \texttt{Type A}. A log-parabola spectral distribution and a point-like source spatial distribution were assumed, and the best parameters were obtained through a forward-folding fitting. For \texttt{Event types}, the datasets for \texttt{Type M}, \texttt{Type A}, and \texttt{Type B} were fit in a joint likelihood approach. Run-by-run corrections to observation conditions were used and are better described in section~\ref{sec:systematics}.

\begin{figure}
    \centering
    \includegraphics[width=0.98\linewidth]{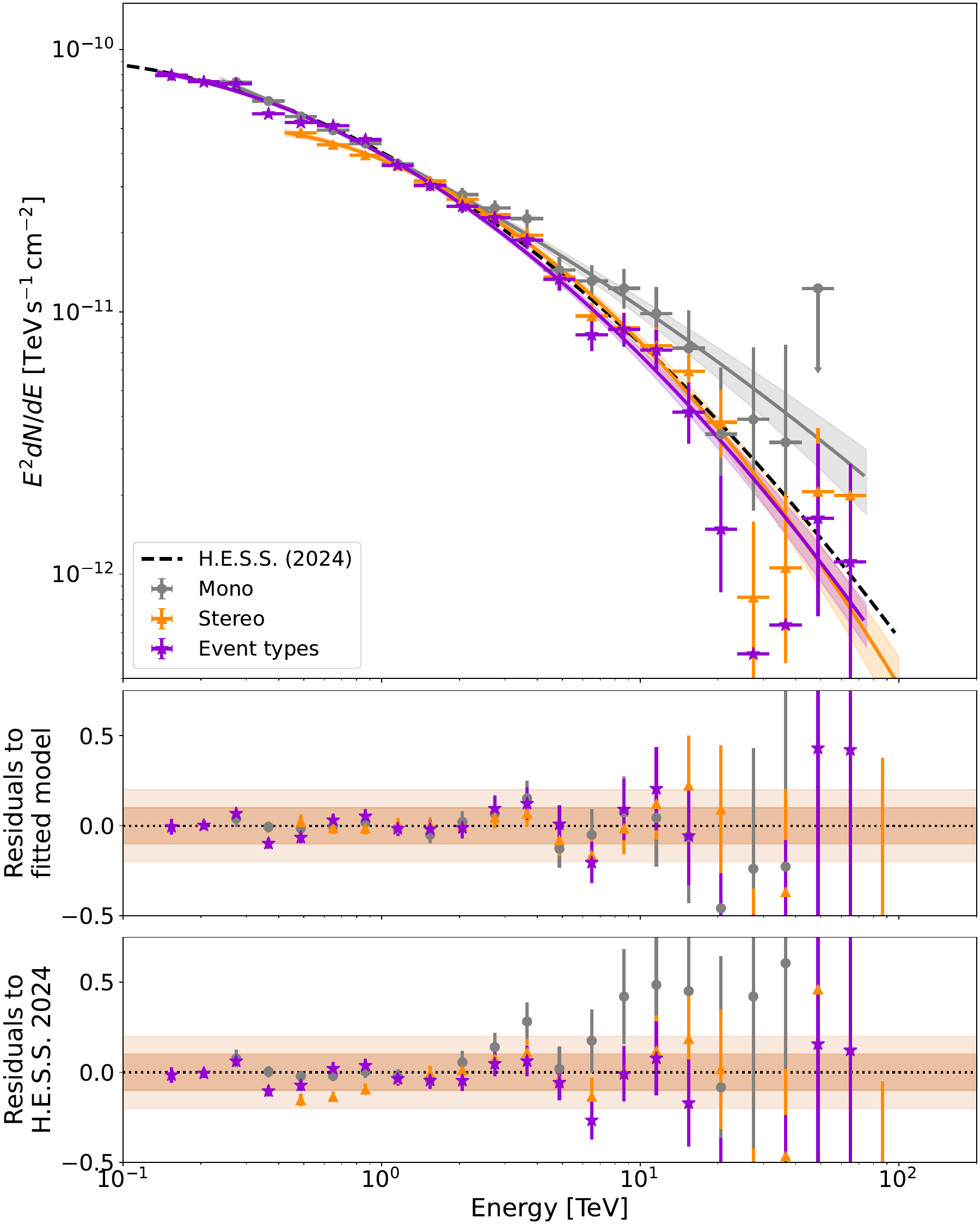}
    \caption{Reconstructed spectral energy distribution of the Crab nebula for the standard \texttt{HAP} configurations, \texttt{Mono} (gray lines and full circles), and \texttt{Stereo} (orange lines and full squares), and \texttt{Event types} (purple lines and stars). The full lines show the best-fit model, and the shaded areas show their statistical uncertainty ranges. The dashed black line shows the best-fit model from~\cite{HessCrab} as a comparison. The center and bottom panels show the residual to the best-fit model of each configuration and to the model of~\cite{HessCrab}, respectively. The dark and light brown bands show the $\pm 10\%$ and $\pm 20\%$ ranges. Run-by-run corrections to observation conditions are applied to all the spectra.}    
    \label{fig:spectrum}
\end{figure}

Figure~\ref{fig:spectrum} shows the resulting spectra for \texttt{Event types} and for the two standard \texttt{HAP} analyses, \texttt{Mono} and \texttt{Stereo}, compared to the best-fit model obtained in~\cite{HessCrab}. The spectra agree well, and residuals to the~\cite{HessCrab} model do not exceed $\pm 10\%$ for the relevant energy ranges, which is within the expected $20\%$ systematic uncertainty range for the flux~\citep{HessCrab2006,HESS:2012bek,Zaborov:2016jub}. The many improvements of the analysis we propose here are clearly observed. Most importantly, a single analysis is able to exploit the whole energy range of H.E.S.S., in contrast to \texttt{Mono} and \texttt{Stereo}, which present a reduced coverage. Furthermore, a much finer spectral reconstruction is obtained, with reduced statistical uncertainties in the model parameters (as seen in the width of the shaded bands). More robustness is also achieved. For \texttt{Mono}, the CT5 reconstruction becomes strongly biased for $E \gtrsim 2 \, \rm{TeV}$ because the camera pixels saturate, leading to artifacts in the energy spectrum. For \texttt{Stereo}, the low-energy end of the spectrum deviates from the expected spectrum. The whole energy range is affected by the forward-folding method, and this leads to a significantly stronger curvature in the final spectrum. Finally, for \texttt{Event types}, a significantly reduced energy threshold (from $\sim 230 \, \rm{GeV}$ to $\sim 130 \, \rm{GeV}$, with two additional energy bins) is also obtained through the additional improvements in the monoscopic configurations.

\section{Systematic uncertainties due to observation conditions}
\label{sec:systematics}

As defined in this work, the event type classification relies on the Hillas image size for each event. The presence of aerosols in the atmosphere and/or telescope mirror degradation can lead to less light being detected by the camera, and thus, to smaller image sizes. Similarly, the night-sky background and changes in the hardware calibration directly affect the final Hillas parameters. The effects can be relatively homogeneous for the whole array (e.g., caused by atmospheric differences) or significantly different per telescope (e.g., caused by mirror degradation and hardware changes). Due to these effects, different observation conditions will lead to a different fraction of events in each type for each energy. This adds another level of complexity to the instrument response functions (IRFs), and if these effects are not corrected for, the final reconstructed fluxes may be subject to strong systematic uncertainties. In \texttt{Event types}, the individual effective areas change rapidly at the boundaries between each type, and thus, this effect is more significant than in the classical \texttt{Mono} and \texttt{Stereo} configurations. For this reason, we used a preliminary version of a run-by-run correction scheme that will be presented in an upcoming publication. The scheme is based on two correction factors: (i) A throughput correction factor that is independent for each telescope and takes the optical efficiency of each telescope into account and is derived from measurements of muon rings~\citep{ThesisAlison}, and (ii) an atmospheric correction factor that reflects the transparency of the atmosphere to Cherenkov photons, derived from the system trigger rate of the CT1-4 telescopes in a procedure similar to that described in~\cite{2014APh....54...25H}.

\begin{figure*}
    \centering
    \includegraphics[width=0.92\linewidth]{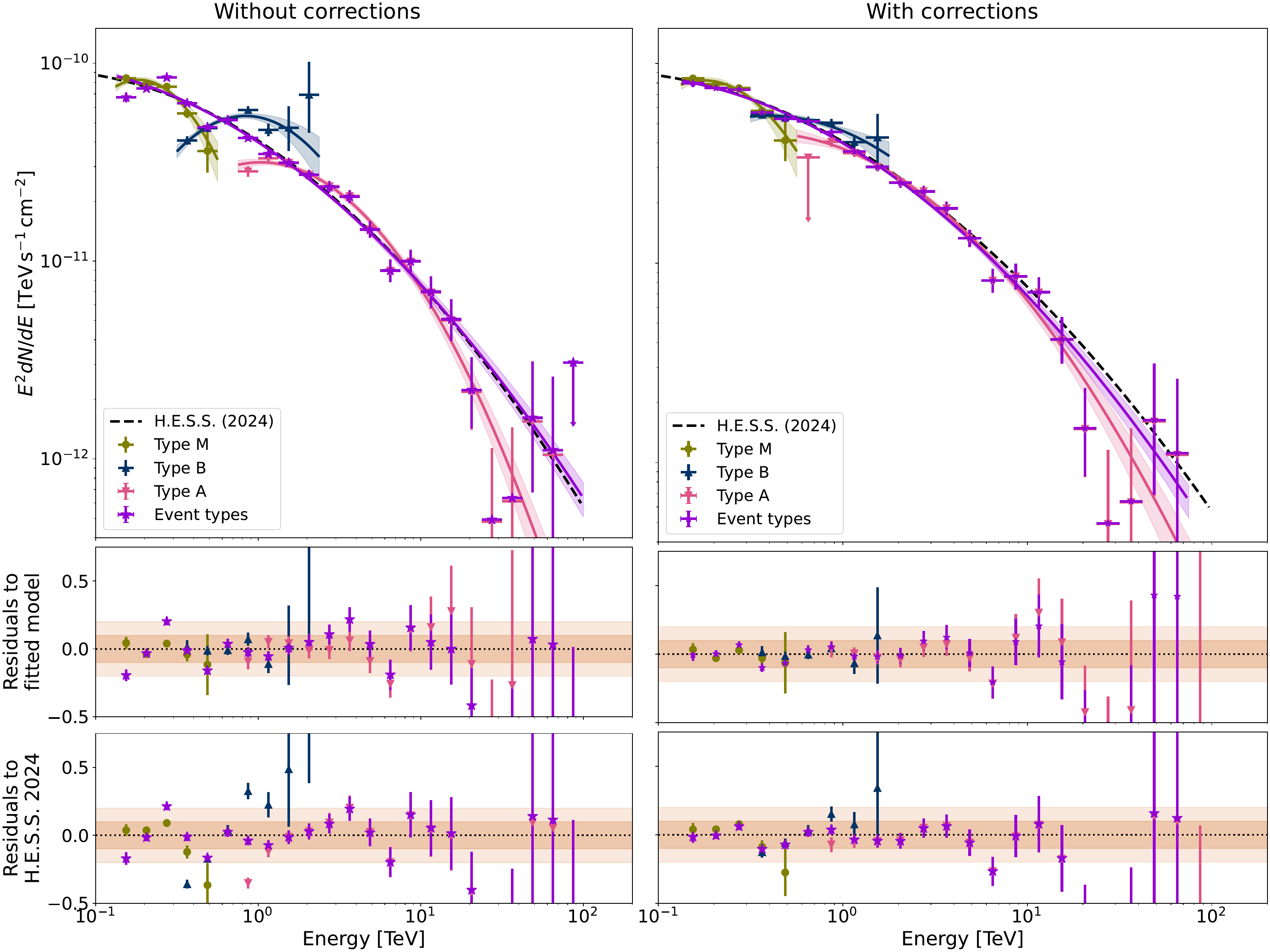}
    \caption{Reconstructed spectral energy distribution of the Crab nebula without (left) and with (right) run-by-run correction for the observation conditions. We show spectra for each individual type, \texttt{Type M} (green lines with full circles), \texttt{Type B} (blue lines with full upward-pointing triangles), and \texttt{Type A} (magenta lines with full downward-pointing triangles), as well as the combined spectrum, \texttt{Event types} (purple lines with stars). The full lines show the best-fit model, and the shaded areas show their statistical uncertainty ranges. The dashed black line shows the best-fit model from~\cite{HessCrab} as a comparison. The center and bottom panels show the residual to the best-fit model of each configuration and to the model of~\cite{HessCrab}, respectively. The dark and light brown bands show the $\pm 10\%$ and $\pm 20\%$ ranges.}    
    \label{fig:sys}
\end{figure*}

Figure~\ref{fig:sys} compares the spectrum with and without run-by-run corrections for the observation conditions. The spectra for individual types are shown to illustrate the effects better, even though they are not expected to be used individually. When no correction is applied, a clear deviation from the expected log-parabola-like spectrum is seen, especially around the transition energies. In the transition between \texttt{Type M} and \texttt{Type B}, a significant overprediction of the former is seen, together with a significant underprediction of the latter. This shows that the chosen observation runs present different observation conditions than the simulated runs that were used in building the IRFs. In the simulations, more events with $E \sim 300 \, \rm{GeV}$ produce usable images in one of the small telescopes compared to real data. More real events than expected therefore end up in \texttt{Type M} rather than \texttt{Type B}, resulting in the deviations shown in the spectrum. When run-by-run corrections are applied, the combined spectrum is stable, even though small deviations are seen in the individual spectra. This highlights the need for a run-by-run correction to observation conditions when event types are used. For bright sources, such as the Crab nebula, these systematic deviations are stronger. For fainter sources, on the other hand, the relative statistical uncertainties are much larger and dominate these deviations.

\section{Conclusions}
\label{sec:conclusions}

We presented the development of an event type analysis in the context of H.E.S.S. The monoscopic and stereoscopic trigger and reconstruction strategies required a definition based on pre-reconstruction parameters, different from the usual definition used in Fermi-LAT. Three event types (\texttt{Type M}, \texttt{Type B}, and \texttt{Type A}) were therefore defined using individual Hillas parameters, as shown in Table~\ref{tab:cuts}. In addition, new improvements were introduced, including the monoscopic improvements presented in~\cite{Unbehaun:2025skr}, the time-based cleaning for CT5 presented in~\cite{Celic:2025wyp}, and further improvements to the stereoscopic configurations, which enable a robust combination of CT5 data with data from small telescopes with improved gamma-hadron separation.

The performance of the new analysis configurations was estimated and compared to the standard \texttt{HAP} configurations, \texttt{Mono} and \texttt{Stereo}. An effective area that successfully covers the whole energy range was found for \texttt{Event types}, in contrast to \texttt{Mono} and \texttt{Stereo}, which focus on energies below and above $E \sim 150 \, \rm{GeV}$, respectively. The improved strategies for gamma-hadron separation led to a higher signal retention and stronger background suppression. An improved energy threshold was obtained due to a size-dependent BDT cut~\citep{Unbehaun:2025skr} and time-based image cleaning~\citep{Celic:2025wyp}. For the monoscopic configurations (\texttt{Mono++} and \texttt{Type M}), significantly improved angular and energy resolutions were found. For the angular reconstruction, the main improvement comes from an NN designed to reconstruct the orientation of the image~\citep{Unbehaun:2025skr}. For the energy resolution, on the other hand, the main improvement comes from time-sensitive images, which can contribute to breaking the degeneracy between the energy and impact parameter~\citep{Unbehaun:2025skr}.

The combined impact on the performance can be seen in the differential sensitivities. When compared to the standard \texttt{HAP} analyses, \texttt{Event types} provides an enhanced sensitivity in the whole energy range and presents improvements of $25-45\%$ in the energy range between 300 GeV and 5 TeV. For the lowest energies ($<100 \, \rm{GeV}$), even larger improvements are found, with a significantly reduced energy threshold.

Intermediate configurations (\texttt{Mono++} and \texttt{Stereo++}) with the old definition (either fully monoscopic or fully stereoscopic), but including new improvements, were also investigated to distinguish the effects of each new development. \texttt{Event types} not only provides a single analysis capable of exploring the whole energy range of H.E.S.S., but also presents improved efficiencies and resolution with respect to \texttt{Stereo++}, especially regarding gamma-hadron separation. This is reflected in a significantly improved sensitivity and arises from the asymmetry in event quality between \texttt{Type B} and \texttt{Type A}, which allows a combined fit to neglect low-quality events. This in turn enhances the contribution of high-quality events.

The proposed analysis was validated with real data by performing a spectral analysis of the Crab nebula using a subset of the data used in~\cite{HessCrab} and \cite{Unbehaun:2025skr}. The expected systematic uncertainties agree with previous results, which shows that the new developments are robust and did not introduce new biases. The potential of \texttt{Event types} can be observed even in a bright source such as the Crab nebula. The full coverage of the energy range was obtained, in contrast to \texttt{Mono} and \texttt{Stereo}, combined with a significantly reduced energy threshold (from $\sim 230 \, \rm{GeV}$ to $\sim 130 \, \rm{GeV}$, adding two energy bins). An improved spectral reconstruction was achieved with reduced statistical uncertainties in the spectral parameters. Furthermore, a more robust spectrum was reconstructed, which compensates for a spurious spectral curvature that was found when only \texttt{Mono} or \texttt{Stereo} were used. Even stronger effects from using event types are expected for fainter sources because of the improved sensitivity and significantly reduced energy threshold.

We also investigated the systematic uncertainties resulting from the varying observation conditions. As the types are defined using Hillas parameters, observation conditions are expected to change the type fraction as a function of energy, leading to spectra with enhanced biases. A run-by-run correction was shown to overcome this and to be a requirement when using \texttt{Event types} analyses.

With this, we presented an analysis capable of exploiting the whole energy range of H.E.S.S. and the application to data of an IACT analysis capable of combining significantly different telescope types with a significantly different energy range coverage. This analysis can be used not only in future H.E.S.S. data, but also might be able to unveil new results in previous data by achieving significantly lower energy thresholds and improvements in the intermediate energies that are equivalent to almost doubling the observation time. The demonstrated effectiveness of the method reinforces its potential in the future CTAO, which will also rely on different telescope types for different energy ranges.

\begin{acknowledgements}
We thank the H.E.S.S. Collaboration for providing simulated data, common analysis tools, and valuable comments on this work. In particular, we thank Werner Hofmann for enlightening discussions on the combined IRFs and sensitivity. We also thank the H.E.S.S. Collaboration for allowing us to use the data on the Crab Nebula in this publication. This research made use of the \textsc{Astropy} (\url{https://www.astropy.org}; \citealt{Astropy2013,Astropy2018,Astropy2022}), \textsc{matplotlib} (\url{https://matplotlib.org}; \citealt{Hunter2007}), \textsc{iminuit} (\url{https://iminuit.readthedocs.io}; \citealt{Dembinski2020}) and \textsc{gammapy} (\url{https://gammapy.org/}; \citealt{gammapy:2023, gammapy:zenodo-1.1}) software packages.
\end{acknowledgements}

\bibliographystyle{aa}
\bibliography{references}

\begin{appendix}
\newpage
\onecolumn

\section{gamma-hadron separation for the stereoscopic configurations}
\label{app:BDT}
For the stereoscopic configurations, gamma-hadron separation was obtained using a BDT with the following variables (the newly introduced variables are denoted by a *):

\begin{itemize}
    \item \textbf{*MSCW14:} Mean-scaled width for CT1-4 considering expectations for gamma-ray events of a given reconstructed energy;
    \item \textbf{*MSCL14:} Mean-scaled length for CT1-4 considering expectations for gamma-ray events of a given reconstructed energy;
    \item \textbf{*MSCWO14:} Mean-scaled width for CT1-4 considering expectations for background events of a given reconstructed energy;
    \item \textbf{*MSCLO14:} Mean-scaled length for CT1-4 considering expectations for background events of a given reconstructed energy;
    \item \textbf{*Width5:} Hillas width for CT5;
    \item \textbf{*Length5:} Hillas length for CT5;
    \item \textbf{*Kurtosis5:} Hillas kurtosis for CT5;
    \item \textbf{*AbsSkewness5:} Absolute value of the Hillas skewness for CT5;
    \item \textbf{*LogDensity5:} $\log_{10}$ of Hillas size / (Hillas width $\times$ Hillas length) for CT5;
    \item \textbf{*LengthOverLogSize5:} Hillas length divided by $\log_{10}$ of Hillas size for CT5;
    \item \textbf{HmaxOverCosZen:} Reconstruted maximum of depth of the air shower corrected by the cosine of the zenith angle;
    \item \textbf{dEoverE:} Spread of the energy reconstructed by each individual telescope;
    
\end{itemize}

Gamma efficiency, $\gamma_{\rm{eff}}$, is defined as the fraction of gamma rays left after gamma-hadron separation and depends strongly on the applied BDT cut. A very stable separation performance was found for a large range of gamma efficiencies. For that reason, a $\gamma_{\rm{eff}} \ge 60\% \, (85\%)$ requirement was introduced for \texttt{Type B} (\texttt{Type A}), ensuring minimal systematic uncertainties due to possible data-MC mismatches.

\section{Sensitivities for 1D W-statistics statistics}
\label{app:sens}

\begin{figure}[h!]
    \centering
    \includegraphics[width=0.65\linewidth]{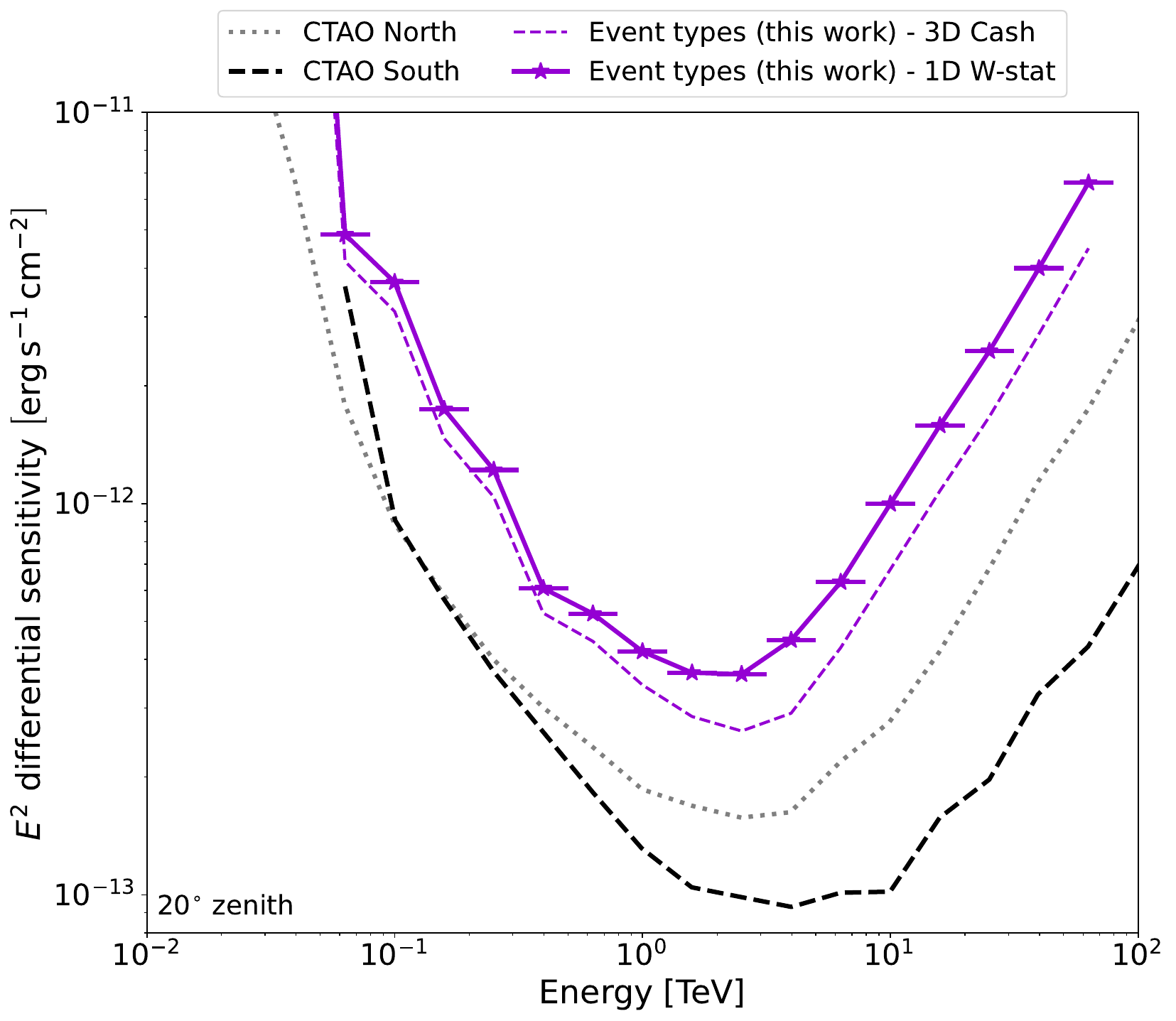}
    \caption{Differential sensitivity as a function of energy for a representative zenith angle of $20^{\circ}$. The sensitivities for \texttt{Event types} using 3D Cash statistics (dashed purple line) and 1D W-statistics (full purple line with stars) are compared to the projected sensitivities for CTAO North (dotted gray line) and CTAO South (full black line)~\citep{cherenkov_telescope_array_observatory_2021_5499840}.}    
    \label{fig:senswstat}
\end{figure}

Figure~\ref{fig:senswstat} shows the sensitivities for \texttt{Event types} if 1D W-statistics are used instead of 3D Cash statistics. Current projected sensitivities for CTAO North and CTAO South~\citep{cherenkov_telescope_array_observatory_2021_5499840} are shown for comparison.

\section{Variability of the improvements over the field of view}
\label{app:fov}

\begin{figure}[h!]
    \centering
    \includegraphics[width=0.65\linewidth]{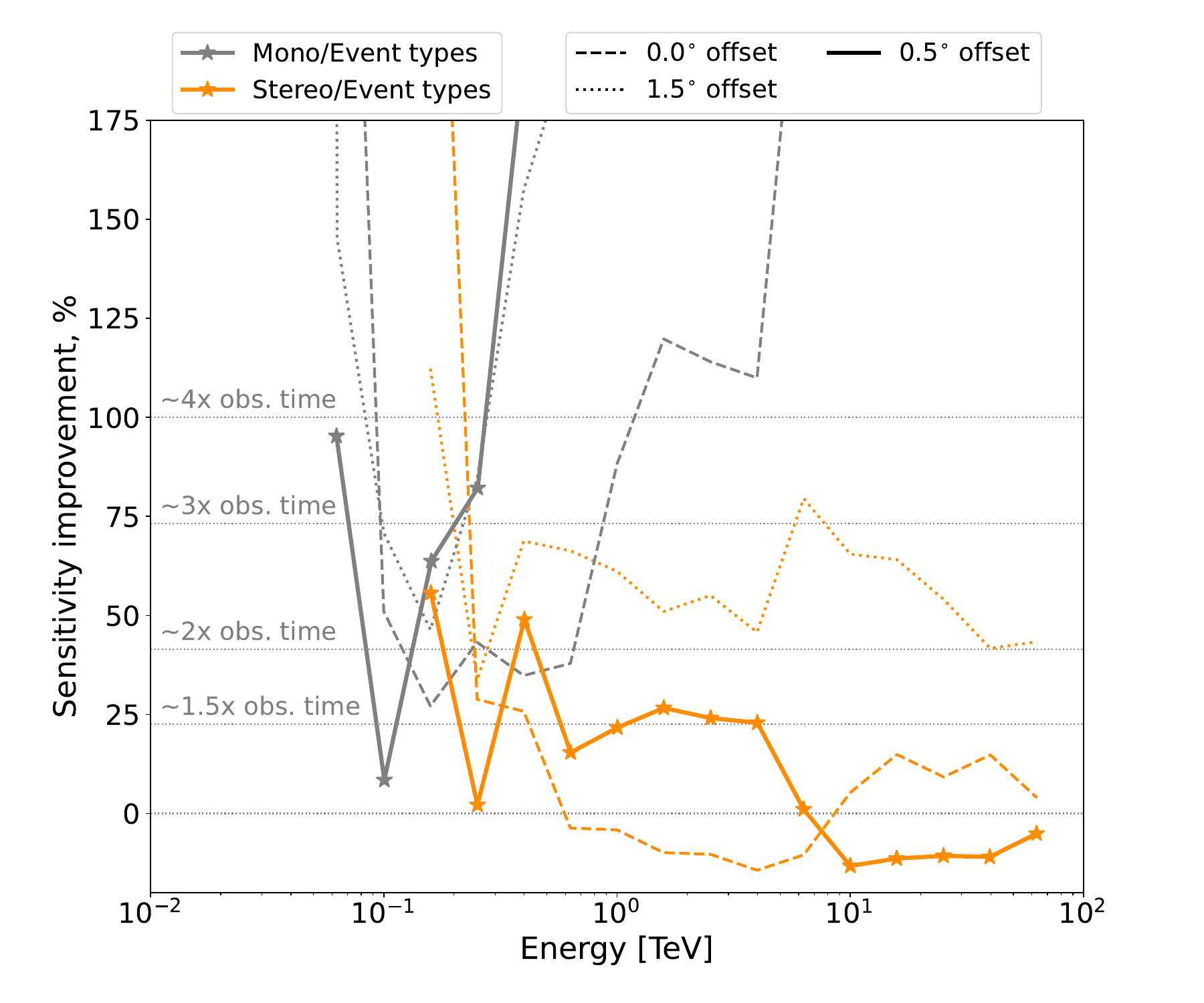}
    \caption{Sensitivity improvement for different observation offsets (defined as the angular distance between the expected source position and the pointing direction. Gray and orange lines show the relative improvements for \texttt{Event types} with respect to \texttt{Mono} and \texttt{Stereo}, respectively.}    
    \label{fig:fov}
\end{figure}

In Section~\ref{sec:sens}, the sensitivity improvement was estimated considering an observation offset of $0.5^{\circ}$. Figure~\ref{fig:fov} shows the improvement for if $0.0^{\circ}$ or $1.5^{\circ}$ are considered. Here, it's noteworthy that the gamma-hadron separation was optimized for $0.5^{\deg}$ offset and, therefore, different levels of improvement for different offsets were expected, as is seen.

\newpage
\newpage

\section{Data points for the differential sensitivities of \texttt{Event types}}
\label{app:data}

Table~\ref{tab:data} lists the data points for the sensitivities of \texttt{Event types} for 1D W-statistics and 3D Cash statistics.

\begin{table}[ht!]
    \centering
        \caption{Data points for the differential sensitivity of \texttt{Event types}.}
    \label{tab:data}
    
\begin{tabular}{c|c|c||c|c}
\hline
\hline

Energy & Min. energy & Max. energy Energy & 3D Cash stat. $E^{2}$ differential sens. & 1D W-stat $E^{2}$ differential sens. \\

$\rm\left[TeV\right]$ & $\rm\left[TeV\right]$ & $\rm\left[TeV\right]$ & $\left[10^{-12} \, \rm{erg \, cm^{-2} \, s^{-1}}\right]$ & $\left[10^{-12} \, \rm{erg \, cm^{-2} \, s^{-1}}\right]$ \\
\hline
\hline
0.0398 & 0.0316 & 0.0501 & 267.2178 & 350.6057 \\
0.0631 & 0.0501 & 0.0794 & 4.1549 & 4.8686 \\
0.1000 & 0.0794 & 0.1259 & 3.0965 & 3.6778 \\
0.1585 & 0.1259 & 0.1995 & 1.4665 & 1.7426 \\
0.2512 & 0.1995 & 0.3162 & 1.0389 & 1.2187 \\
0.3981 & 0.3162 & 0.5012 & 0.5247 & 0.6073 \\
0.6310 & 0.5012 & 0.7943 & 0.4446 & 0.5232 \\
1.0000 & 0.7943 & 1.2589 & 0.3440 & 0.4188 \\
1.5849 & 1.2589 & 1.9953 & 0.2856 & 0.3697 \\
2.5119 & 1.9953 & 3.1623 & 0.2623 & 0.3664 \\
3.9811 & 3.1623 & 5.0119 & 0.2913 & 0.4478 \\
6.3096 & 5.0119 & 7.9433 & 0.4284 & 0.6300 \\
10.0000 & 7.9433 & 12.5893 & 0.6788 & 0.9982 \\
15.8489 & 12.5893 & 19.9526 & 1.0768 & 1.5835 \\
25.1189 & 19.9526 & 31.6228 & 1.6681 & 2.4531 \\
39.8107 & 31.6228 & 50.1187 & 2.7150 & - \\
63.0957 & 50.1187 & 79.4328 & 4.4898 & -  \\
\hline
\hline
\end{tabular}

\end{table}

\end{appendix}

\end{document}